# Phonons in SrTiO$_3$ analyzed by difference bond-length spectrum

Wilfried Wunderlich,
Tokai University, Graduate School of Engineering, Material Science Department,
259-1292 Kanagawa-ken, Hiratsuka-shi, Kitakaname 1117, Japan
Email: wi-wunder@rocketmail.com, Phone: +81-90-7436-0253

Phonons change remarkable the interatomic bond length in solids and this work suggest a novel method how this behavior can be displayed and analyzed. The bond-length spectrum is plotted for each of the different atomic bonding types. When comparing the bond-length to an un-deformed crystal by the so-called difference bond length spectrum, the effect of phonons is clearly visible. The Perovskite lattice of SrTiO$_3$ is used as an example and several lattice vibration modes are applied in a frozen phonon calculation in a 2x2x2 supercell. Ab-initio DFT simulations using the Vasp software were performed to calculate the density of states. The results show the important finding reported here first, that certain phonon interactions with shorter Ti-O bonds decrease the band gap, while changes in the Sr-Ti bond length enlarge the band gap.

**Introduction**
Research on phonon has recently become more popular, not only because they influence thermal properties but mainly because electron-phonon coupling has a great influence on transport properties. Experimental results on semi-conducting Nb-doped SrTiO$_3$ [1] show, that this can be described by the relaxation time as an important parameter, besides carrier concentration and effective mass [2,3], for understanding thermoelectric phenomena. Due to the same phenomena also doped NaTaO3 is suggested as a new thermoelectric material [4]. Electron-phonon coupling causes nonlinearities in frequency resolved reflectivity [5,6] and permittivity [7]. The phonon spectrum of any material, as e.g. the perovskite SrTiO3 [8] and CaTiO$_3$ [9], is calculated by the Hellmann-Feynman forces in an extended supercell (2x2x2) by the frozen phonon method, in which direct displacements on the atoms are applied [9]. The systematic search for new materials requires further understanding of the electron-phonon interaction and this paper describes two new features. Usually in phonon calculations simple modes are modeled in order to calculate the phonon spectrum. However, the idea described as main topic in this paper, is that at high temperatures not only one single phonon mode is present but several phonon modes, interacting with each other. In other word, heavy vibrations of the lattice brings atoms closer together, which are usual apparent. Secondly, for analyzing this behavior, a novel display method for analyzing is introduced, the difference bond length spectrum.

**Computation details**
The frozen phonon method [9] was applied on 2x2x2 supercells of SrTiO3 as shown in fig.1a. The perovskite lattice with spacegroup Pm3m (221) has five atomic positions, Sr, Ti, O(1), O(2), O(3), where the Oxygen positions are structurally independent, but they are symmetry equivalent. It is however, well-known, that these cations can trigger vibrations among each other and a shift of a part of the whole Oxygen sublattice is likely to occur. Hence, three possible Oxygen modes, just O(1), O(1) and O(2) together or all three O were assumed structural models, in the following named as O-1, O-2 and O-3. For each group of atoms (Sr, Ti, O-1, O-2 and O-3), all together five different sorts, the mixed phonon modes were applied, as explained in the following. As shown in fig 1b) a coherent sinusoidal vibration, c) complete uniaxial shift, d) coherent opposing vibrations, e) vibration and shift, f) vibration and shift in another mode were applied, all together also five modes. The change in atomic coordinates of each of the eight supercells are marked wit $a_i$, i=1..8 and referred to in fig. 1b. The value for the elongation was x=0.05, a value less than 10% of the lattice parameter usually tolerated in frozen phonon calculations. This algorithm was applied in [100] direction, and additionally also in [110] and [111] direction. For the vibrational mode in fig. 1b) the displacements in the three directions are displayed in the ball-stick mode, in views along y- and z- zone axis. The bright atoms are Sr, blue Ti and red Oxygen, and bonds less than 0.24nm are displayed, usual the TiO$_6$- octahedron bonds. Fig 2 shows that in some cases additional bonds fall below this limit, especially when all Oxygens are displaced.
The bond spectrum is the frequency sum of all bonds which lengths fall within a certain length interval, as displayed in fig 3a for the undistorted $SrTiO_3$ lattice. In single atomic metals is function is usually called pair-correlation function (PCF) or radial distribution function (RDF). The $TiO_6$-octahedron bonds with 0.19nm length lead for the 2x2x2 supercell to a sum of 96 bonds, where the bonds to atoms the rim of the supercell need to be included. For obtaining the information, which atomic type forms for example the 0.27nm bonds, the atom type resolved spectra were plotted (fig, 3b). The result is: All 0.27nm bonds are Sr-O bonds in total 350 bonds. Finally, the longest bonds 0.3905nm, the same as the lattice constant, are formed between Sr-atoms. In the following, the phonon mode fig. 1 b is applied as shown in fig. 3 b and c.

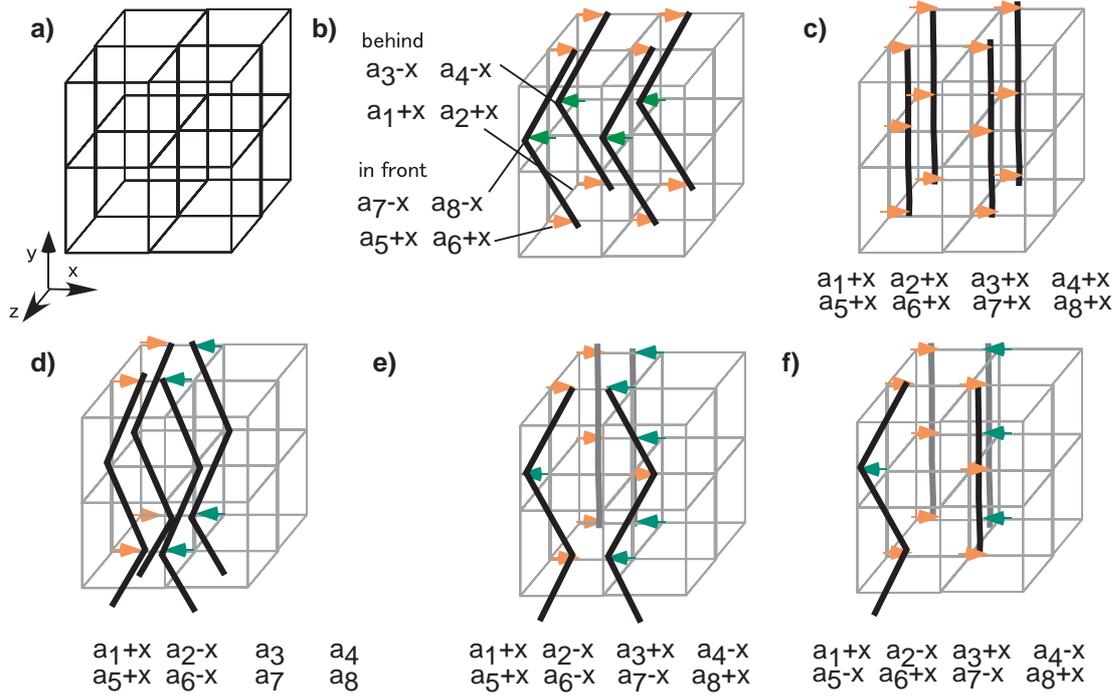

Fig. 1 Phonon modes and their related atomic coordinate transformations applied to each of one atoms sort in SrTiO3 in this study: a) undeformed 2x2x2 supercell, b) coherent sinusoidal vibration, c) complete uniaxial shift, d) coherent opposing vibrations, e) vibration and shift, f) vibration and shift in another mode.

This paper suggests a novel characterization method for phonons as shown in fig. 3, based on the interatomic bonding length. The effect of the phonon becomes clearly visible, by plotting the difference of the bonding length compared to the undistorted SrTiO3 lattice. The differences appear as negative intensities, where the original bond length is absent, and instead on its shorter or longer bond (left- and right-hand side) counts of new bond length appear. This is a typical pattern for a LO or TA phonon. For comparison other pattern in the difference bond length spectrum are displayed in fig. 4: Shear deformation leads to several shorter or longer bonds than the undeformed case (fig. 4a) Isotropic lattice expansion or uniaxial deformation (fig. 4 b, c) leads to a pair of negative and positive peaks. One displaced O atom effects already many bonds (fig. 4d). The Slater and Last phonon modes [7] lead to difference phonon spectra shown in fig. 4 e and f. As in the Slater mode both, Ti and O, are shifted towards to each other, a very short Ti-O bond appears. The examples show, that this analysis will find a wide range of applications, for example in analyzing the output atomic coordinates of Molecular Dynamics simulations.

All the above mentioned 5 frozen phonon modes were applied to the 5 atom sorts in the supercell in 3 directions of elongation, leading in total 5x5x3=75 structure models. For each of them ab-intio calculations based on the density-functional theory (DTF) using the VASP version 4.5 software [10] were applied. After plotting the density of states, also the band-structure was plotted. The experimental obtained band-gap for $SrTiO_3$ is 3.2eV, the VASP results are smaller (2eV), but this has no significant influence to the following interpretation and can be corrected by using the LDA+U self-energy correction or calibration assuming a threshold value [11], as done in this paper. The effective band mass was calculated as described in [3,12] from the curvature of bands near the CB minimum and VB maximum at the $\Gamma$-point.

**Results and Discussion**

The plot of the density of states (DOS) is shown in fig 5., where the rows refer to the distortion mode (fig. 1), the columns to the direction of elongation [100],

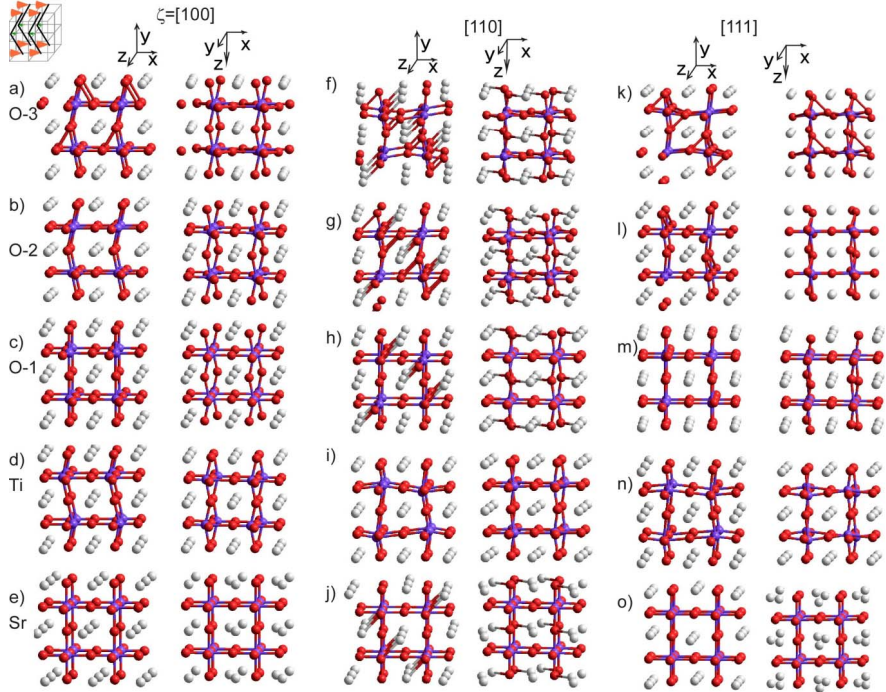

Fig. 2 Phonon mode from fig 1 b) applied to different atoms of the SrTiO3 lattice. In each display the left picture shows the projection near-[001] zone axis and the right picture along [010] zone axis. a) all three O-atom types displaced, b) two O atom types displaced, c) only the O(1) type displaced, d) Ti displaced, e) Sr displaced; f)-g) same as a) to e) but instead of shift in (100) direction, shifted in (110) direction, k)-o) same as a)-e), but shifted in (111)-direction.

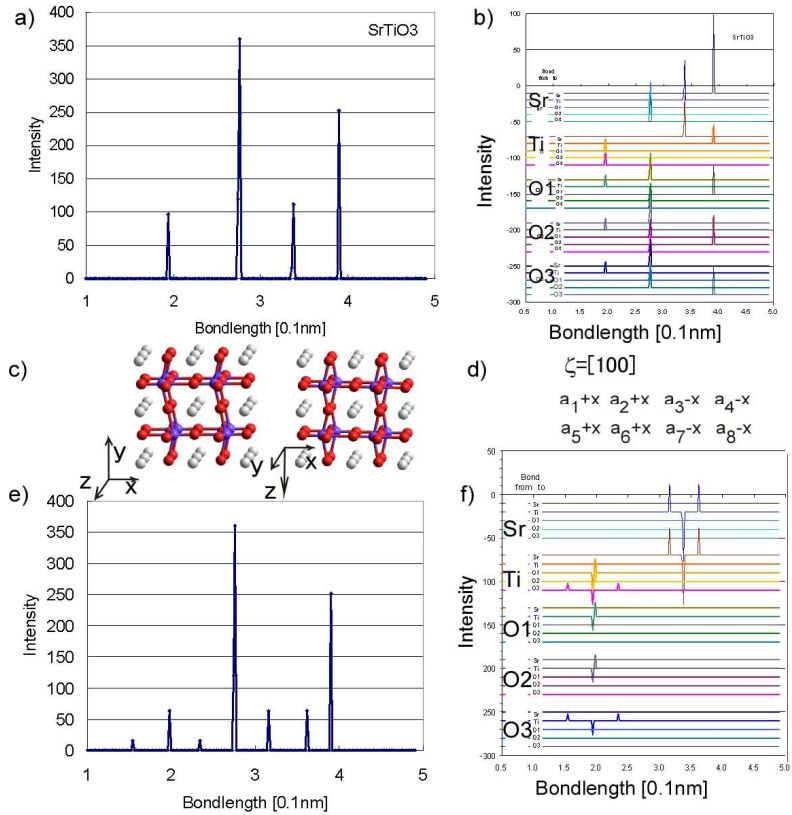

Fig. 3 Bondlength spectra (Intensity of the interatomic bonding length) for a) undistorted SrTiO3, and e) for the phonon mode defined in d) and shown in c). This mode corresponds to the one in fig. 1b. Fig. 3 b) shows the bondlength spectra resolved for each bonding type in the order Sr-Sr, Sr-Ti, Sr-O1, Sr-O2, Sr-O3, Ti-Sr, Ti-Ti etc. Fig. 3 f) shows the difference bondlength spectra of this phonon mode compared to undistorted SrTiO$_3$ (fig, 2b).

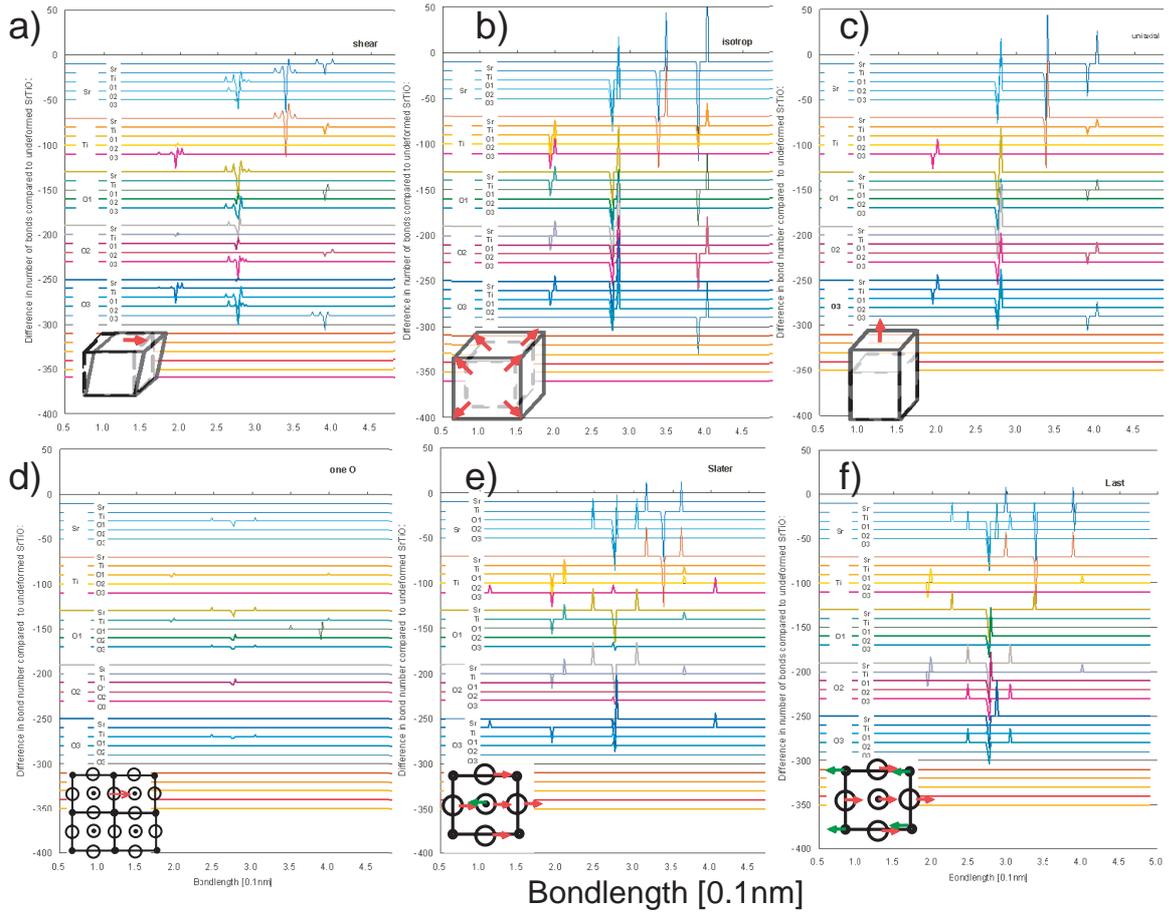

Fig. 4 Difference bondlength spectra for different deformation and phonon modes, a) shear deformation, b) isotropic expansion, c) uniaxial tension, d) shift of one Oxygen atom, e) Slater phonon mode, f) Last phonon mode.

[110], or [111] and in each of the sub-pictures the order refers to O-3 (O(1),O(2),O(3)), O-2 (O(1),O(2)) O-1 (O(1)), Ti-, Sr- atoms distorted and at the bottom undistorted $SrTiO_3$ for comparison. In many cases the band width remains unchanged, but there are also many cases where the band-gap became smaller, as marked with dark, filled dot. The few case where the band-gap became larger are marked with an open circle. The check using the difference bond length displays showed that a shortening of the Ti-O bonds will reduce the band gap, while a change in the Sr-Ti bond length widens the band gap, with a few exceptions, when both conditions are fulfilled. The important finding that the bandgap is drastically reduced simple due to phonon vibrations is reported in this paper for the first time.

The detailed analysis of the forth case in fig. 5 d is shown in fig. 6. For clarification the sketch of the vibration type (fig. 6a), the atomic positions (fig. 6b) and the difference bond length spectrum (fig. 6c) are displayed. The electronic bandstructure is shown for the important directions Γ-X and Γ-M, others do not play an important role for properties of perrovskite [3]. The electronic bandstructure is shown for different magnitudes of elongation, ξ=a[100], with a=0.05 (fig. 6e), 0.04 (f), 0.03 (g) and 0.01 (h). The band gap becomes smaller with increasing displacement and finally CB and VB are overlapped. This means a metallic behavior for high temperatures and indeed in experiments a drastic reduction of the resistance of perovskites is observed [1,2,4]. The effective mass deduced form the band curvature of the lowest conduction band remains unchanged compared to undeformed $SrTiO_3$.

These frozen phonon calculation results show in this paper for the first time that certain phonon modes in $SrTiO_3$ can reduced the band-structure properties in such a way that electric conduction becomes possible. The reason is the interaction of mixed phonon modes consisting of different displacements in different crystallographic directions as shown in this paper for the first time. In reality all phonon modes are present at all temperatures, each with amplitudes that depend in their own way on temperature. It is a challenge for the future to clarify these dependences and compare these static calculations with the actual temperature dependence of the band gap width and the relaxation time. The suggested difference bond length spectrum will be a useful tool towards reaching the goal understanding the electron-phonon interaction.

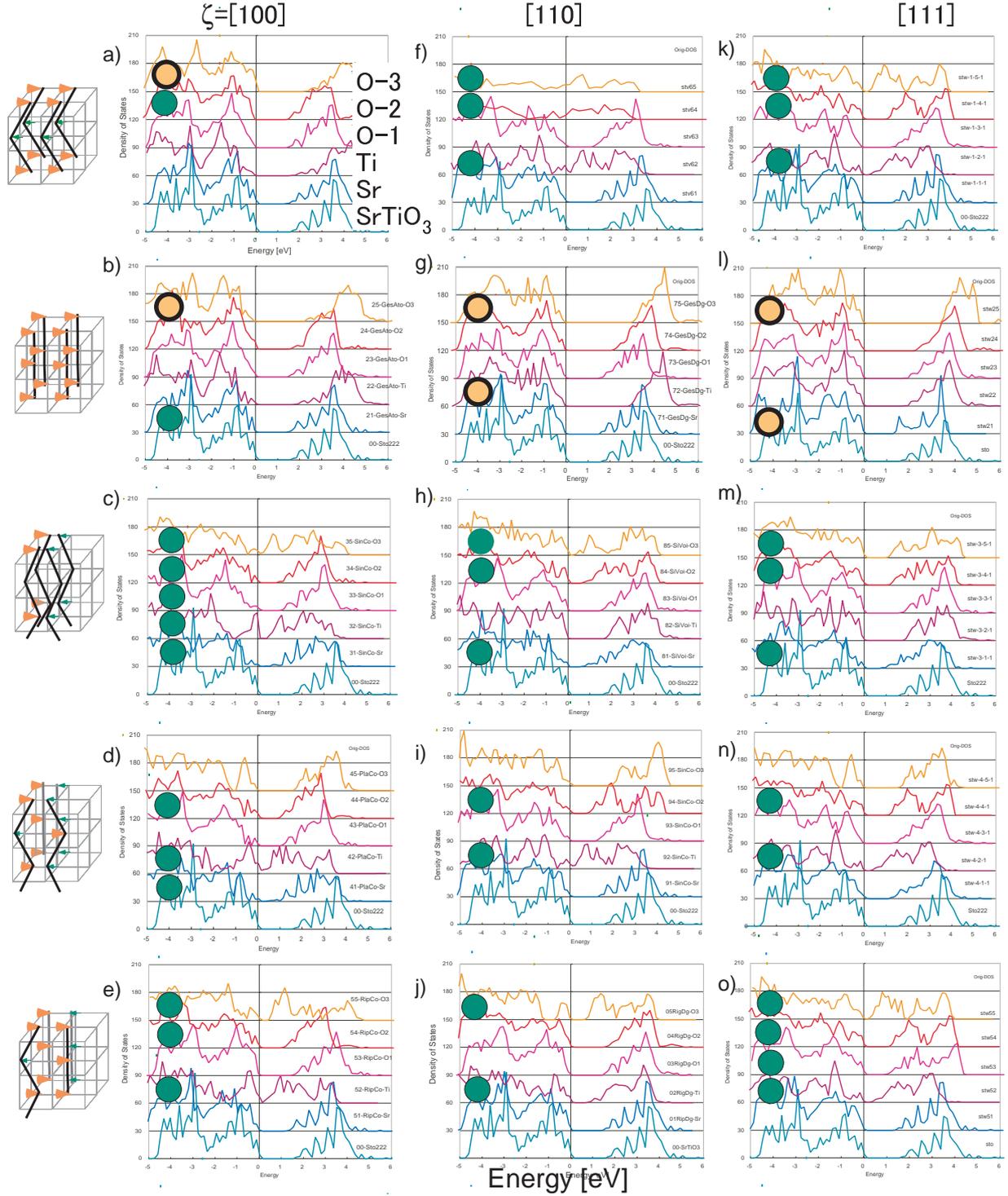

Fig. 5. Density of states as a function of the Energy (Fermi level is at $E_F=0$) for all phonon modes calculated in this study. a), f), k) sinusoidal mode with the same order of atomic displacement (O3, O2, O1, Ti, Sr) as in fig 3 including the lower DOS of undeformed $SrTiO_3$ as comparison. Displacements in b), g), l) are uniaxial shift displacements corresponding to fig. 1b), c) h) m) correspond to fig. 1c) and so on. The rows a)- e) refer to displacements along $\zeta=[100]$, f)- j) along $\zeta=[110]$, k)-o) along $\zeta=[111]$, respectively. The open orange circle are cases where the bandgap is remarkably larger than undeformed $SrTiO_3$, the closed green circles are cases where the bandgap is remarkably smaller than undeformed $SrTiO_3$.

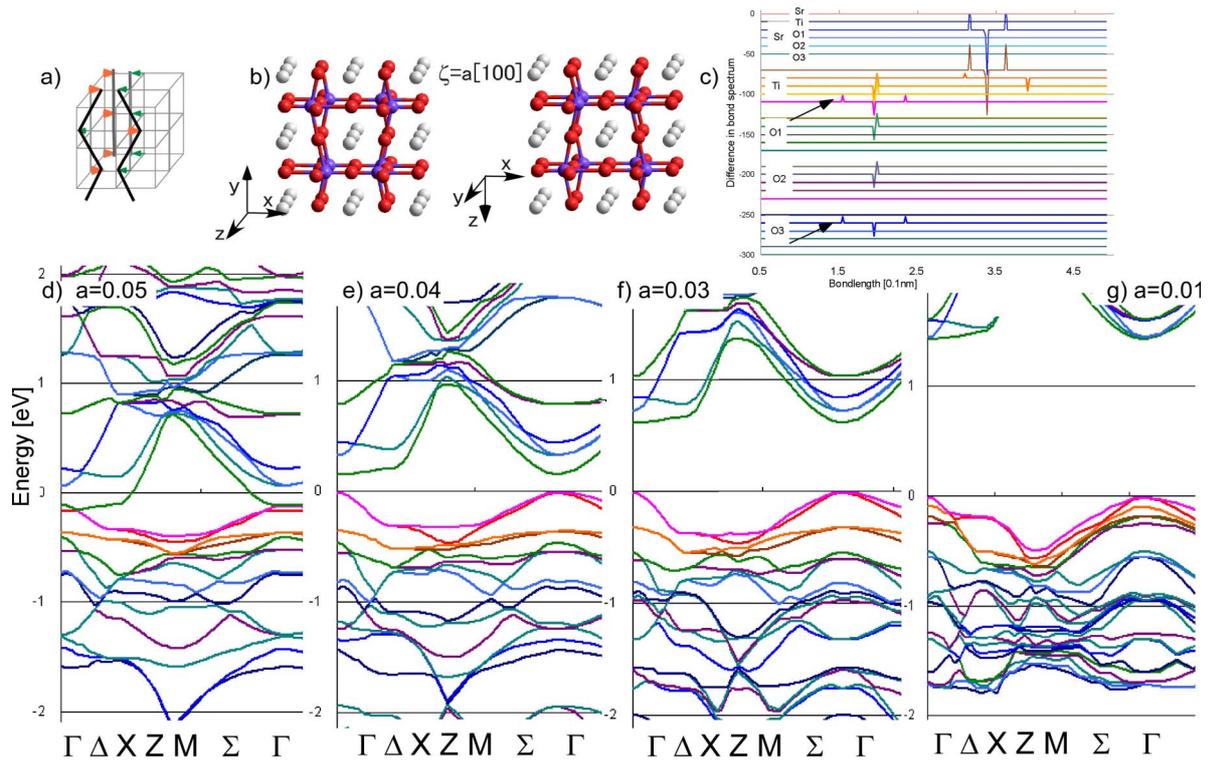

Fig. 6 a) Phonon mode of fig. 1d applied on Ti-atoms in ξ= a[100] (same as in fig. 4d, 4.th from top), b) its atomic coordinates, c) its difference bond length spectrum with shortened Ti-O Bonds marked with arrows, leading to d)-g) bandstructures, in which the bandgap shrinks with increasing distortion amplitude g) a=0.01, f) a=0.03, e) a=0.03, d) a=0.05 as marked.

**Conclusions**

Frozen phonon calculations with atomic distortion modes for all atomic types in several directions were performed. The results are:

(i) The plot of the density of states and the electronic bandstructure showed in this paper for the first time that some mixed phonon modes reduce the band gap, others leave it unchanged and others widen it. The shift of the electronic bandgap occurs linearly and the effective mass is unchanged.

(ii) The difference bond length spectrum analysis showed that a shortening of Ti-O bonds is responsible for a smaller bandgap, and that of Sr-Ti for a wider bandgap.